\documentclass[pdflatex,sn-vancouver,Numbered]{sn-jnl}

\usepackage{graphicx}%
\usepackage{multirow}%
\usepackage{amsmath,amssymb,amsfonts}%
\usepackage{amsthm}
\usepackage{mathrsfs}%
\usepackage[title]{appendix}%
\usepackage{xcolor}%
\usepackage{textcomp}%
\usepackage{manyfoot}%
\usepackage{booktabs}%
\usepackage{algorithm}%
\usepackage{algorithmic}%
\usepackage{listings}%
\usepackage{soul}
\renewcommand{\hl}[1]{#1}
\soulregister\cite7
\soulregister\citep7
\soulregister\eqref7
\soulregister\ref7
\usepackage{adjustbox}
\usepackage{diagbox}
\usepackage{bm}

\makeatletter
\let\orcidlogo\@undefined
\makeatother
\usepackage{orcidlink}

\usepackage{float}
\usepackage{chngcntr}

\usepackage{siunitx}

\usepackage{dcolumn}

\newtheorem{remark}{Remark}%

\begin{document}

\title[Article Title]{Structural perspectives from quantum states and measurements in optimal state discrimination}


\author[1]{\fnm{Hyunho} \sur{Cha} \orcidlink{0009-0008-2933-6989}}\email{aiden132435@cml.snu.ac.kr}

\author*[1]{\fnm{Jungwoo} \sur{Lee} \orcidlink{0000-0002-6804-980X}}\email{junglee@snu.ac.kr}

\affil[1]{\orgdiv{NextQuantum and Department of Electrical and Computer Engineering}, \orgname{Seoul National University}, \orgaddress{\city{Seoul}, \postcode{08826}, \country{Republic of Korea}}}


\abstract{Quantum state discrimination
refers to a class of techniques to identify a specific quantum state through a \textit{positive operator-valued measure}. In this work, we investigate how structural information
can influence our ability to determine or bound the optimal discrimination probability. \hl{First, as background, we note that for single-qubit pure-state ensembles, pairwise fidelities determine the optimal discrimination probability, whereas in higher dimensions they do not in general.
As an illustrative application of this observation, we give a closed-form fidelity-based reformulation of the optimal discrimination probability for three equiprobable single-qubit states with equal pairwise fidelities.
Secondly, we show that the information of measurement operators that vanish in the optimal solution can be used to refine upper bounds on the optimal discrimination probability, often yielding tighter bounds.}
}




\maketitle

\subsection*{Acknowledgments}
This work is in part supported by the National Research Foundation of Korea (NRF, RS-2024-00451435 (20\%), RS-2024-00413957 (20\%)), Institute of Information \& communications Technology Planning \& Evaluation (IITP, RS-2021-II212068 (10\%), RS-2025-02305453 (15\%), RS-2025-02273157 (15\%), RS-2025-25442149 (10\%) RS-2021-II211343 (10\%)) grant funded by the Ministry of Science and ICT (MSIT), Institute of New Media and Communications (INMAC), and the BK21 FOUR program of the Education, Artificial Intelligence Graduate School Program (Seoul National University), and Research Program for Future ICT Pioneers, Seoul National University in 2025.

\section{Introduction}
\label{sec:pure_state_discrimination}

Quantum state discrimination (QSD) stands as a cornerstone within quantum information theory, essential for quantum measurements of which the output is used to identify quantum states \citep{chefles2000quantum, barnett2009quantum, bergou2010discrimination, bae2015quantum}. The task involves constructing a \textit{positive operator-valued measure} (POVM), which is characterized by a set of \textit{positive semidefinite} (PSD) operators that sum to the identity operator \citep{davies1970operational, peres1990neumark}.


\hl{QSD has developed into a broad subject that includes general theoretical frameworks \citep{helstrom1969quantum, holevo1973statistical, chefles2000quantum, barnett2009quantum, bae2013structure, bae2015quantum, weir2017optimal}, optimality criteria and minimax formulations \citep{yuen2003optimum, d2005minimax, barnett2008conditions, nakahira2015generalized}, and structured families such as symmetric and geometrically uniform states \citep{ban1997optimum, barnett2001minimum, eldar2002quantum, eldar2004optimal}. Experimental demonstrations approaching the Helstrom limit have also been reported \citep{barnett1997experimental, clarke2001experimental, higgins2009mixed, solis2017experimental, jagannathan2022demonstration}.}

When the \textit{a priori} probabilities of states are unknown, the minimax QSD approach can be applied to ensure the least possible maximum error across all states \citep{d2005minimax}. In contrast, when the \textit{a priori} probabilities are available, the \textit{pretty good measurement} (PGM), also known as the \textit{square root measurement}, is recognized as one of the central strategies \citep{hausladen1994pretty, hausladen1996classical}. Among several criteria for optimality \citep{helstrom1969quantum, davies1978information}, it optimizes the \textit{average} success probability. This analytical tool has been extensively studied and is leveraged to infer the classical labels of quantum states selected from an ensemble.

We denote by $n$ the number of qubits. Let the states lie in a $d$-dimensional Hilbert space $\mathcal{H}_d$, where $d = 2^n$. Consider the discrimination of $N$ pure states $\{ | \psi_i \rangle \}_{i = 1}^N$ with $| \psi_i \rangle$ occurring with \textit{a priori} probability $p_i$. This is described as the following ensemble:
\begin{eqnarray*}
\mathcal{E} = \{ | \Tilde{\psi}_i \rangle \}_{i = 1}^N = \{ \sqrt{p_i} | \psi_i \rangle \}_{i = 1}^N.
\end{eqnarray*}
The state $| \psi_i \rangle$ is a normalized version of $| \Tilde{\psi}_i \rangle$.

The QSD problem involves finding the POVM $E = \{ E_i \}_{i = 1}^N$ that maximizes the probability of successfully determining the correct state. Formally, we must solve the following optimization problem:
\begin{eqnarray}
\label{eqn:optimization_formulation}
\underset{E}{\text{max}} & \quad & P_{\mathcal{E}, E}^\text{success} \nonumber \\
\text{subject to} & \quad & E_i \succeq 0 \quad \forall i, \\
&& \sum_{i = 1}^N E_i = I, \nonumber
\end{eqnarray}
where
\begin{eqnarray*}
P_{\mathcal{E}, E}^\text{success} \equiv \sum_{i = 1}^N w(\mathcal{E},E)_i = \sum_{i = 1}^N \text{tr}(| \Tilde{\psi}_i \rangle \langle \Tilde{\psi}_i | E_i) = \sum_{i = 1}^N p_i \text{tr}(| \psi_i \rangle \langle \psi_i | E_i)
\end{eqnarray*}
is the success probability and $I$ is the identity operator on $\mathcal{H}_d$. The constraints in Eq.~\eqref{eqn:optimization_formulation} enforce that all measurement operators in $E$ must be positive semidefinite and that they must sum to the identity operator.

The global optimal solution of Eq.~\eqref{eqn:optimization_formulation} can be computed with arbitrary precision. In fact, the problem is an instance of semidefinite programming (SDP), which is a convex optimization problem. Let $\hat{E}(\mathcal{E}) = \{ \hat{E}_i (\mathcal{E}) \}_{i = 1}^N$ and $P_\mathcal{E}^\text{opt} = P_{\mathcal{E}, \hat{E}}^\text{success}$ denote the optimal POVM and the optimal success probability for $\mathcal{E}$, respectively. Also, we denote \(w(\mathcal{E},\hat{E}(\mathcal{E}))\) by \(\hat{w}(\mathcal{E})\). One notable property of $P_\mathcal{E}^\text{opt}$ is its invariance under unitary transformations $| \Tilde{\psi}_i \rangle \mapsto U | \Tilde{\psi}_i \rangle$.

In this article, we explore two structural properties of the QSD problem. \hl{We first revisit the single-qubit pure-state setting, where pairwise fidelities determine the optimal discrimination probability.} In higher dimensions, however, pairwise fidelities \hl{do not in general suffice}. \hl{As an illustrative application of this observation, we give a closed-form fidelity-based expression of the optimal discrimination probability for three equiprobable single-qubit states with equal pairwise fidelities.} \hl{We then investigate how the knowledge of measurement operators that vanish in the optimal solution can be used to refine upper bounds on the optimal success probability.} \hl{We also provide examples of subsets and supersets of nonvanishing operators that can be identified without SDP}, suggesting a potential avenue for reducing computational costs.

\section{\hl{Pretty Good Measurement}}

PGM is an alternative approach to SDP that offers a balance between performance and efficiency \citep{hausladen1994pretty, hausladen1996classical}. The measurement operators are defined as
\begin{equation}
\label{equation:PGM_E_i_definition}
E_i \equiv S^{-1/2} | \Tilde{\psi}_i \rangle \langle \Tilde{\psi}_i | S^{-1/2}
\end{equation}
where
\begin{eqnarray*}
S \equiv \sum_{i = 1}^{N} | \Tilde{\psi}_i \rangle \langle \Tilde{\psi}_i |
\end{eqnarray*}
is the density matrix of the ensemble and the inverse is the Moore-Penrose pseudoinverse \citep{penrose1955generalized}. Intuitively, the term $S^{-1/2}$ effectively whitens the states to some extent, making them closer to being orthogonal and thus more distinguishable when measured.

Although PGM does not in general achieve the optimal success probability, its performance falls within the following range \citep{barnum2002reversing, montanaro2007distinguishability}:
\begin{eqnarray}
\label{equation:pgm_square_bound}
( P_\mathcal{E}^\text{opt} )^2 \leq P_\mathcal{E}^{\text{PGM}} \leq P_\mathcal{E}^\text{opt}.
\end{eqnarray}

\section{\hl{Motivation:} Sufficiency of pairwise fidelities for single-qubit \hl{pure} states}
\label{sec:sufficiency}

\hl{We provide backgrounds and establish notation. Much of the discussion in this section follows directly from the standard geometry of the Bloch sphere.}

Suppose we are not given $\mathcal{E}$ itself, but we know the Gram matrix $G$ of pairwise unnormalized inner products:
\begin{eqnarray*}
G_{ij} = \langle \Tilde{\psi}_i | \Tilde{\psi}_j \rangle.
\end{eqnarray*}
\hl{The Gram matrix $G$ determines the family $\mathcal{E}$ on its span up to isometry \citep{akibue2019perfect}. Since in our setting the states are regarded inside a fixed $\mathcal{H}_d$, this isometry can be extended to a unitary on $\mathcal{H}_d$. Hence unitary-invariant quantities such as $\hat{w}(\mathcal{E})$ and $P_\mathcal{E}^\text{opt}$ are determined by $G$.}

Now consider the matrix $F$ of pairwise unnormalized fidelities:
\begin{eqnarray*}
F_{ij} = | \langle \Tilde{\psi}_i | \Tilde{\psi}_j \rangle |.
\end{eqnarray*}
The normalized fidelities are
\begin{eqnarray*}
\hat{F}_{ij} = | \langle \psi_i | \psi_j \rangle | = \frac{| \langle \Tilde{\psi}_i | \Tilde{\psi}_j \rangle |}{\sqrt{p_i p_j}} = \frac{F_{ij}}{\sqrt{F_{ii} F_{jj}}}.
\end{eqnarray*}
We would like to know whether $\hat{w}(\mathcal{E})$ and $P_\mathcal{E}^\text{opt}$ can be uniquely determined even when we lack the phase information in $G$ and only have access to $F$. The answer is affirmative, but only if $d = 2$. To see this, we use the generalized Bloch vector \citep{kimura2003bloch, kimura2005bloch, bengtsson2017geometry}.

First, define a set of $d^2$ projection operators $\pi_{jk} = | j \rangle \langle k |$, where $j, k \in \{ 0, 1, \cdots , d - 1 \}$ denote the computational basis states. Then define three types of operators and their union:
\begin{align*}
\alpha_j & \equiv \sqrt{\frac{2}{j(j + 1)}} \left( 
\sum_{k = 0}^{j} \pi_{kk} - (j + 1) \pi_{j + 1, j + 1} \right), \\
\beta_{jk} & \equiv \pi_{jk} + \pi_{kj}, \quad j < k, \\
\gamma_{jk} & \equiv i(\pi_{jk} - \pi_{kj}), \quad j < k, \\
\{ \lambda_i \} & \equiv \{ \alpha_j \} \cup \{ \beta_{jk} \} \cup \{ \gamma_{jk} \}.
\end{align*}
The set $\{ \lambda_i \}$ consists of $d^2 - 1$ operators, and they satisfy
\begin{eqnarray}
\label{eq:lambda_trace}
\text{tr}(\lambda_i) = 0 \quad \text{and} \quad \text{tr}(\lambda_i \lambda_j) = 2\delta_{ij}.
\end{eqnarray}
The fidelity between two states can be represented in terms of their corresponding Bloch vectors. For a $d$-dimensional pure state $\rho$, its (generalized) Bloch vector $\mathbf{r}^{(\rho)} \in \mathbb{R}^{d^2 - 1}$ is defined such that
\begin{eqnarray}
\label{eq:bloch_definition}
\rho = \frac{1}{d} \left( I + \frac{d(d - 1)}{2} \lambda \cdot \mathbf{r}^{(\rho)} \right),
\end{eqnarray}
where $|| \mathbf{r}^{(\rho)} || = 1$. Using Eqs.~\eqref{eq:lambda_trace} and \eqref{eq:bloch_definition}, the fidelity between two states can be expressed as
\begin{eqnarray*}
\sqrt{\text{tr}(\rho \sigma)} = \sqrt{\frac{1 + (d - 1) \mathbf{r}^{(\rho)} \cdot \mathbf{r}^{(\sigma)}}{d}}.
\end{eqnarray*}
Given $F$ for the pure state ensemble $\mathcal{E}$, we have
\begin{eqnarray}
\label{equation:bloch_vector_inner_product}
\mathbf{r}^{(| \psi_i \rangle \langle \psi_i |)} \cdot \mathbf{r}^{(| \psi_j \rangle \langle \psi_j |)} = \frac{d \cdot F_{ij}^2 / F_{ii} F_{jj} - 1}{d - 1},
\end{eqnarray}
that is, the pairwise inner products of the Bloch vectors are determined by $F$. This implies that $F$ determines the set of Bloch vectors $\left\{ \mathbf{r}^{(| \psi_i \rangle \langle \psi_i |)} \right\}_{i = 1}^N$ up to an orthogonal transformation \citep[Theorem 7.3.11]{horn2012matrix}. We remark that Eq.~\eqref{equation:bloch_vector_inner_product}, combined with the nonnegativity of fidelity, gives an alternative proof that the cosine similarity between two Bloch vectors lies within the range $\left[ -\frac{1}{d - 1}, 1 \right]$, as noted in Refs. \citep{jakobczyk2001geometry, kimura2003bloch}.

Let $\mathcal{E} = \{ | \Tilde{\psi}_i \rangle \}_{i = 1}^N$ and $\mathcal{E}^\prime = \{ | \Tilde{\psi}^\prime_i \rangle \}_{i = 1}^N$ be two different ensembles of single-qubit states for which the matrix of pairwise unnormalized fidelities is $F$. Then there exists an orthogonal matrix $R$ that maps $\left\{ \mathbf{r}^{(| \psi_i \rangle \langle \psi_i |)} \right\}_{i = 1}^N$ to $\left\{ \mathbf{r}^{(| \psi^\prime_i \rangle \langle \psi^\prime_i |)} \right\}_{i = 1}^N$, i.e., $R \mathbf{r}^{(| \psi_i \rangle \langle \psi_i |)} = \mathbf{r}^{(| \psi^\prime_i \rangle \langle \psi^\prime_i |)}$. If $R \in \text{SO}(3)$, it corresponds to a unitary operator on $\mathcal{H}_2$ \citep[p. 175]{nielsen2010quantum}, i.e., there exists a unitary $U$ such that $U | \Tilde{\psi}_i \rangle = | \Tilde{\psi}^\prime_i \rangle$ for all $i$. It follows that the optimal POVMs for $\mathcal{E}$ and $\mathcal{E}^\prime$ are also related by $U$, so $\hat{w}(\mathcal{E}) = \hat{w}(\mathcal{E}^\prime)$ and $P_\mathcal{E}^\text{opt} = P_{\mathcal{E}^\prime}^\text{opt}$. If $R \notin \text{SO}(3)$, it may be decomposed as a proper rotation in $\text{SO}(3)$ followed by a reflection across the $xy$-plane. Using the expansion
\begin{eqnarray*}
\rho = \frac{1}{2} \begin{pmatrix}
1 + \mathbf{r}^{(\rho)}_3 & \mathbf{r}^{(\rho)}_1 - i \mathbf{r}^{(\rho)}_2 \\
\mathbf{r}^{(\rho)}_1 + i \mathbf{r}^{(\rho)}_2 & 1 - \mathbf{r}^{(\rho)}_3 \\
\end{pmatrix},
\end{eqnarray*}
the reflection of a Bloch vector across the $xy$-plane corresponds to
\begin{eqnarray*}
\rho \mapsto X \rho^\top X.
\end{eqnarray*}
Then, the transformation
\begin{eqnarray*}
E_i \mapsto X E_i^\top X \quad \forall i
\end{eqnarray*}
gives the same success probability, so we still have $\hat{w}(\mathcal{E}) = \hat{w}(\mathcal{E}^\prime)$ and $P_\mathcal{E}^\text{opt} = P_{\mathcal{E}^\prime}^\text{opt}$.

For $d > 2$, the optimal success probability is not determined by $F$. This is because not every orthogonal transformation of the Bloch vector can be described as a unitary transformation on $\mathcal{H}_d$ (the converse is true \citep{schlienz1995description}). Indeed, starting with a counterexample for $(d, N) = (2, 3)$, we can readily extend this to higher dimensions and/or larger numbers of states. Developing algorithms based on reduced representations, such as pairwise fidelities for single-qubit states or the Gram matrix for multi-qubit states, is expected to offer greater efficiency compared to SDP.


\hl{To illustrate this observation, we use it to discriminate three equiprobable and equidistant single-qubit states.}
Before proceeding, we remark that the construction of the optimal POVM for this case is already known \citep{ban1997optimum, bae2013minimum}. The purpose of this \hl{example} is to express $P_\mathcal{E}^\text{opt}$ \hl{only} in terms of fidelity, which is apparently simple only for three states.

\hl{In the literature, the condition of being `equidistant' is typically characterized by the Gram matrix \cite{paiva2010quantum, roa2011conclusive, zhang2018unambiguous}:}
\begin{eqnarray*}
\langle \psi_i | \psi_j \rangle = \alpha \in \mathbb{C}, \quad \forall i < j.
\end{eqnarray*}
We relax this condition and only require equal pairwise fidelities:
\begin{eqnarray*}
| \langle \psi_i | \psi_j \rangle | = \alpha \in \mathbb{R}_+, \quad \forall i \neq j.
\end{eqnarray*}
Let
\begin{eqnarray*}
\mathcal{E} = \{ | \Tilde{\psi}_1 \rangle, | \Tilde{\psi}_2 \rangle, | \Tilde{\psi}_3 \rangle \}
\end{eqnarray*}
and suppose we are given that
\begin{eqnarray}
\label{equation:equi_condition}
| \langle \Tilde{\psi_i} | \Tilde{\psi_j} \rangle | = \begin{cases}
1 / 3 & \text{if } i = j \\
\alpha / 3 & \text{if } i \neq j
\end{cases}
\end{eqnarray}
for some real number $\alpha \in [1 / 2, 1]$. From Section~\ref{sec:sufficiency}, this will uniquely determine $P_\mathcal{E}^\text{opt}$. Therefore, it suffices to calculate $P_\mathcal{E}^\text{opt}$ for any $\mathcal{E}$ satisfying Eq.~\eqref{equation:equi_condition}. We consider
\begin{eqnarray*}
| \psi_1 \rangle = | 0 \rangle, \quad | \psi_2 \rangle = \alpha | 0 \rangle + \sqrt{1 - \alpha^2} | 1 \rangle, \quad \text{and} \quad | \psi_3 \rangle = \alpha | 0 \rangle + e^{i \theta} \sqrt{1 - \alpha^2} | 1 \rangle,
\end{eqnarray*}
where $\theta$ is chosen such that $| \langle \psi_2 | \psi_3 \rangle | = \alpha$.

\hl{It can be easily seen that $\mathcal{E}$ is an ensemble of \textit{geometrically uniform states} \citep{eldar2002quantum, eldar2003mixed, eldar2004optimal, bae2015quantum}}, where the unitary that represents the symmetry corresponds to a rotation of the Bloch sphere by $2\pi / 3$ about the axis $\sum_{i = 1}^3 \mathbf{r}^{(| \psi_i \rangle \langle \psi_i |)}$. Also, it is known that the optimal POVM for geometrically uniform states is PGM \citep{ban1997optimum}. Using the explicit formula for the square root of a $2 \times 2$ matrix \citep{levinger1980square}, \hl{$S^{-1 / 2}$ in Eq.~\eqref{equation:PGM_E_i_definition}} is calculated as follows (for simplicity, the constants $1 / 3$ are ignored, as they cancel out):
\begin{gather*}
S = \begin{pmatrix}
1 + 2\alpha^2 & (1 + e^{-i\theta})\alpha\sqrt{1 - \alpha^2} \\
(1 + e^{i\theta})\alpha\sqrt{1 - \alpha^2} & 2(1 - \alpha^2)
\end{pmatrix} = \begin{pmatrix}
A & B \\
C & D
\end{pmatrix}, \\
\tau = A + D = 3, \quad \delta = AD - BC = 3(1 - \alpha^2), \\
s = \sqrt{\delta} = \sqrt{3(1 - \alpha^2)}, \quad t = \sqrt{\tau + 2s} = \sqrt{3 + 2s}, \\
S^{1 / 2} = \frac{1}{t} \begin{pmatrix}
A + s & B \\
C & D + s
\end{pmatrix} = \frac{1}{\sqrt{3 + 2s}} \begin{pmatrix}
A + s & B \\
C & D + s
\end{pmatrix}
\end{gather*}
(for $\delta$, see Appendix~\ref{section:delta_derivation}). From symmetry, the optimal success probability is given by
\begin{figure}[t]
\centering
\includegraphics[width=0.7\textwidth]{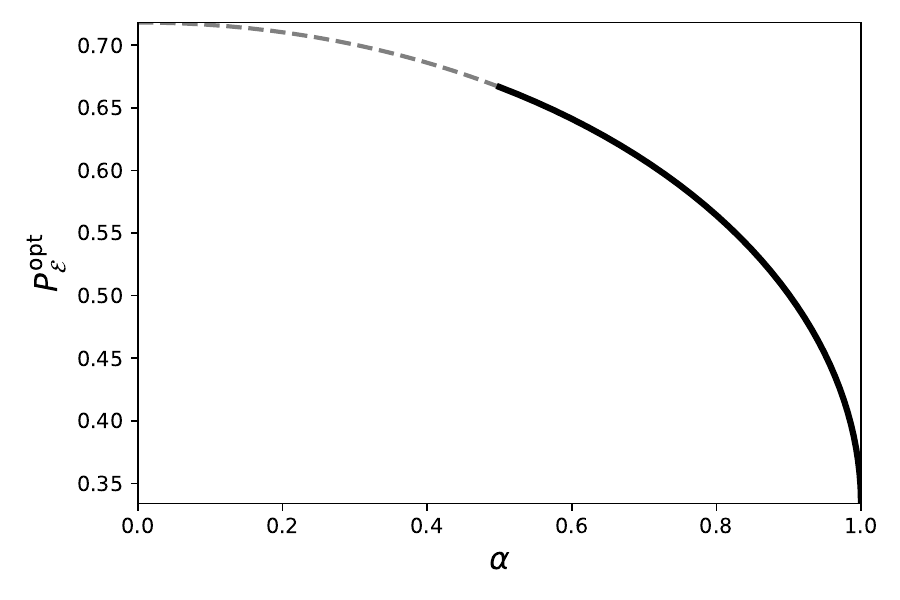}
\caption{\label{fig:equidistant_ellipse} $P_\mathcal{E}^\text{opt}$ vs. $\alpha$ for Eq.~\eqref{equation:equi_condition}. The realizable points lie along the solid line.}
\end{figure}
\begin{align*}
P_\mathcal{E}^\text{opt}
& = \sum_{i = 1}^3 \frac{1}{3} \text{tr}\!\left( | \psi_i \rangle \langle \psi_i | S^{-1 / 2} | \psi_i \rangle \langle \psi_i | S^{-1 / 2} \right) \\
& = \text{tr}\!\left(| \psi_1 \rangle \langle \psi_1 | S^{-1 / 2} | \psi_1 \rangle \langle \psi_1 | S^{-1 / 2}\right) \\
& = \big(\langle \psi_1 | S^{-1 / 2} | \psi_1 \rangle\big)^2 \\
& = \big(S^{-1 / 2}_{11}\big)^2 \\
& = (3 + 2s) \left(\frac{D + s}{(A + s)(D + s) - BC}\right)^2 \\
& \displaybreak[1] \\
& = (3 + 2s) \left(\frac{2s^2 / 3 + s}{s(3 + 2s)}\right)^2 \\
& = \frac{3 + 2s}{9} \\
& = \frac{2\sqrt{3}}{9}\sqrt{1 - \alpha^2} + \frac{1}{3},
\end{align*}
which is an arc of an ellipse (see Fig.~\ref{fig:equidistant_ellipse}).

\section{Sparsity in QSD}
\label{section:sparsity}
\hl{The optimal POVM $\hat{E} (\mathcal{E})$ may be sparse (i.e., include zero measurement operators), meaning that the measurement effectively ignores certain states in the given ensemble \citep{deconinck2010qubit}.} A simple example of this is the family of mirror-symmetric states \citep{andersson2002minimum}. Intuitively, for states with low probabilities, it is more effective to forgo identifying them and instead assign portions of the identity operator to the more probable states.

For generality, we will henceforth use the following notation for an ensemble:
\begin{eqnarray}
\label{dm_notation_ensemble}
\mathcal{E} = \{ \Tilde{\sigma}_i \}_{i = 1}^N = \{ p_i \sigma_i \}_{i = 1}^N,
\end{eqnarray}
where $\sigma_i$ is a unit-trace density matrix (possibly a mixed state). Let $\text{I}_+(\mathcal{E}) \equiv \{ i | \hat{E}_i (\mathcal{E}) \succ 0 \}$ denote the set of indices of positive definite (optimal) measurement operators for $\mathcal{E}$. Obviously, we have $\sum_{i \in \text{I}_+(\mathcal{E})} \hat{E}_i (\mathcal{E}) = I$ and
\begin{eqnarray*}
P_\mathcal{E}^\text{opt} = \sum_{i = 1}^N \text{tr}(| \Tilde{\psi}_i \rangle \langle \Tilde{\psi}_i | \hat{E}_i (\mathcal{E})) = \sum_{i \in \text{I}_+(\mathcal{E})} \text{tr}(| \Tilde{\psi}_i \rangle \langle \Tilde{\psi}_i | \hat{E}_i (\mathcal{E})).
\end{eqnarray*}
Suppose we are given $\text{I}_+(\mathcal{E})$, but not $\hat{E}(\mathcal{E})$ itself. Then the optimization Eq.~\eqref{eqn:optimization_formulation} is equivalent to
\begin{eqnarray}
\label{equation:scaled_optimization}
\underset{E}{\text{max}} & \quad & p_+(\mathcal{E}) P_{\Tilde{\mathcal{E}}, E}^\text{success} \nonumber \\
\text{subject to} & \quad & E_i \succeq 0 \quad \forall i, \\
&& \sum_{i = 1}^{|\text{I}_+(\mathcal{E})|} E_i = I, \nonumber
\end{eqnarray}
where
\begin{eqnarray*}
E = \{ E_i \}_{i = 1}^{|\text{I}_+(\mathcal{E})|}, \quad p_+(\mathcal{E}) \equiv \sum_{i \in \text{I}_+(\mathcal{E})} p_i, \quad \text{and} \quad \Tilde{\mathcal{E}} \equiv \left\{ \frac{\Tilde{\sigma}_i}{p_+(\mathcal{E})} \mid i \in \text{I}_+(\mathcal{E}) \right\}.
\end{eqnarray*}
Note that $\Tilde{\mathcal{E}}$ is a valid ensemble with scaled probabilities. In Eq.~\eqref{equation:scaled_optimization}, if we ignore the constant $p_+(\mathcal{E})$ in the objective, the optimization is precisely in the form of QSD, so it is clear that
\begin{eqnarray}
\label{equation:pruned_relation}
P_\mathcal{E}^\text{opt} = p_+(\mathcal{E}) P_{\Tilde{\mathcal{E}}}^\text{opt}.
\end{eqnarray}
It turns out that this relation allows us to tighten the upper bound on $P_\mathcal{E}^\text{opt}$ in many cases.

\subsection{Bound from PGM}

Instead of Eq.~\eqref{equation:pgm_square_bound}, we start from the following bound, which is shown to be tight \citep{renes2017better}:
\begin{align}
\label{equation:pgm_square_bound_tighter}
P_\mathcal{E}^\text{opt} \leq \sqrt{\frac{N - 1}{N}\left( P_\mathcal{E}^{\text{PGM}} - \frac{1}{N} \right)} + \frac{1}{N}.
\end{align}
Define $P_\mathcal{E}^{\text{PGM}+}$ as the PGM success probability for $\mathcal{E}$ restricted to $\text{I}_+(\mathcal{E})$. That is,
\begin{eqnarray*}
P_\mathcal{E}^{\text{PGM}+} = \sum_{i \in \text{I}_+(\mathcal{E})} \text{tr}(\Tilde{\sigma}_i E_i),
\end{eqnarray*}
where
\begin{eqnarray*}
S \equiv \sum_{i \in \text{I}_+(\mathcal{E})} \Tilde{\sigma}_i \quad \text{and} \quad E_i \equiv \begin{cases}
S^{-1 / 2} \Tilde{\sigma}_i S^{-1 / 2} & \text{if } i \in \text{I}_+(\mathcal{E}) \\
0 & \text{otherwise}
\end{cases}.
\end{eqnarray*}
One can verify that $p_+(\mathcal{E}) P_{\Tilde{\mathcal{E}}}^{\text{PGM}} = P_\mathcal{E}^{\text{PGM}+}$. Applying Eq.~\eqref{equation:pgm_square_bound_tighter} to $\Tilde{\mathcal{E}}$ and using Eq.~\eqref{equation:pruned_relation}, we get
\begin{align}
\label{equation:pgm_improved_bound}
P_\mathcal{E}^\text{opt} & \leq p_+(\mathcal{E}) \left\{ \sqrt{\frac{|\text{I}_+(\mathcal{E})| - 1}{|\text{I}_+(\mathcal{E})|}\left( P_{\Tilde{\mathcal{E}}}^{\text{PGM}} - \frac{1}{|\text{I}_+(\mathcal{E})|} \right)} + \frac{1}{|\text{I}_+(\mathcal{E})|} \right\} \nonumber \\
& = p_+(\mathcal{E}) \left\{ \sqrt{\frac{|\text{I}_+(\mathcal{E})| - 1}{|\text{I}_+(\mathcal{E})|}\left( \frac{P_\mathcal{E}^{\text{PGM}+}}{p_+(\mathcal{E})} - \frac{1}{|\text{I}_+(\mathcal{E})|} \right)} + \frac{1}{|\text{I}_+(\mathcal{E})|} \right\}.
\end{align}
Intuitively, Eq.~\eqref{equation:pgm_improved_bound} is likely tighter than Eq.~\eqref{equation:pgm_square_bound_tighter}, as the latter includes states that do not contribute to $P_\mathcal{E}^\text{opt}$. Examples and counterexamples follow.

\subsubsection{Mirror-symmetric states}

\hl{As the optimal POVM for three mirror-symmetric states is already known \citep{andersson2002minimum, samsonov2009minimum, bae2015quantum}, this section and Section~\ref{sec:mirror_bound_2} are not intended to solve that family anew. Rather, we use it as an analytically tractable benchmark for comparing the original bound with the refined bound.}

\begin{figure}[t]
\centering
\includegraphics[width=0.8\textwidth]{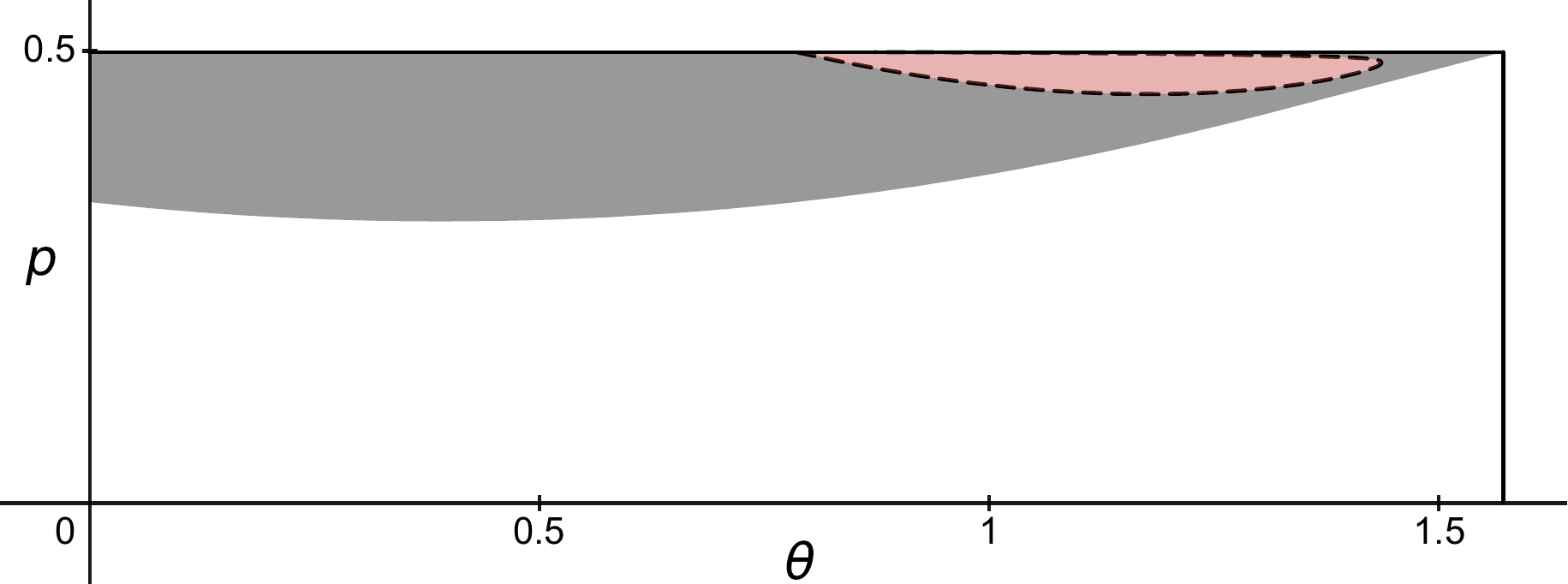}
\caption{\label{fig:mirror_improve_region} The black region represents points where both Eq.~\eqref{equation:p_theta_inequality} and Eq.~\eqref{equation:desired_inequality} are satisfied, whereas the red region represents points where Eq.~\eqref{equation:p_theta_inequality} holds but Eq.~\eqref{equation:desired_inequality} does not.}
\end{figure}

Following the notations in Eq.~\eqref{dm_notation_ensemble}, mirror-symmetric states are defined as
\begin{gather}
N = 3, \quad p_1 = p_2 = p, \quad p_3 = 1 - 2p, \quad \sigma_i = | \psi_i \rangle \langle \psi_i |, \nonumber \\
| \psi_1 \rangle = \cos\theta | 0 \rangle + \sin\theta | 1 \rangle, \quad | \psi_2 \rangle = \cos\theta | 0 \rangle - \sin\theta | 1 \rangle, \quad | \psi_3 \rangle = | 0 \rangle,\label{equation:mirror_definition}
\end{gather}
where $0 \leq p \leq 1 / 2$ and $0 \leq \theta \leq \pi / 2$. Then the POVM $\{ E_1, E_2, E_3 \}$ in Eq.~\eqref{equation:PGM_E_i_definition} is calculated as follows:
\begin{gather*}
S = \sum_{i = 1}^{3} p_i \sigma_i = \begin{pmatrix}
2p\cos^2 \theta + 1 - 2p & 0 \\
0 & 2p\sin^2 \theta
\end{pmatrix} = \begin{pmatrix}
1 - 2p\sin^2 \theta & 0 \\
0 & 2p\sin^2 \theta
\end{pmatrix}, \\
S^{-1 / 2} = \begin{pmatrix}
\frac{1}{\sqrt{1 - 2p\sin^2 \theta}} & 0 \\
0 & \frac{1}{\sqrt{2p\sin^2 \theta}}
\end{pmatrix}, \\
E_1 = S^{-1 / 2} p\sigma_1 S^{-1 / 2} = \begin{pmatrix}
\frac{p\cos^2 \theta}{1 - 2p\sin^2 \theta} & \frac{\sqrt{p} \cos\theta}{\sqrt{2(1 - 2p\sin^2 \theta)}} \\
\frac{\sqrt{p} \cos\theta}{\sqrt{2(1 - 2p\sin^2 \theta)}} & \frac{1}{2}
\end{pmatrix}, \\
E_2 = S^{-1 / 2} p\sigma_2 S^{-1 / 2} = \begin{pmatrix}
\frac{p\cos^2 \theta}{1 - 2p\sin^2 \theta} & \frac{-\sqrt{p} \cos\theta}{\sqrt{2(1 - 2p\sin^2 \theta)}} \\
\frac{-\sqrt{p} \cos\theta}{\sqrt{2(1 - 2p\sin^2 \theta)}} & \frac{1}{2}
\end{pmatrix}, \\
E_3 = S^{-1 / 2} (1 - 2p)\sigma_3 S^{-1 / 2} = \begin{pmatrix}
\frac{1 - 2p}{1 - 2p\sin^2 \theta} & 0 \\
0 & 0
\end{pmatrix}.
\end{gather*}
Therefore,
\begin{align*}
P_\mathcal{E}^{\text{PGM}} & = \sum_{i = 1}^{3} p_i \text{tr}(\sigma_i E_i) \\
& = 2p \left( \frac{p\cos^4 \theta}{1 - 2p\sin^2 \theta} + \frac{\sqrt{2p} \cos^2 \theta \sin\theta}{\sqrt{1 - 2p\sin^2 \theta}} + \frac{\sin^2 \theta}{2} \right) + \frac{(1 - 2p)^2}{1 - 2p\sin^2 \theta} \\
& = 2p \left( \frac{\sqrt{p} \cos^2 \theta}{\sqrt{1 - 2p \sin^2 \theta}} + \frac{\sin\theta}{\sqrt{2}} \right)^2 + \frac{(1 - 2p)^2}{1 - 2p\sin^2 \theta}
\end{align*}
Meanwhile, for large enough $p$, it was shown that the optimal POVM distinguishes only between $\sigma_1$ and $\sigma_2$. Specifically, if
\begin{eqnarray}
\label{equation:p_theta_inequality}
p \geq 1 / \{ 2 + \cos\theta (\cos\theta + \sin\theta) \},
\end{eqnarray}
then $\text{I}_+(\mathcal{E}) = \{ 1, 2 \}$ and $p_+(\mathcal{E}) = 2p$. In this case, $P_\mathcal{E}^{\text{PGM}+}$ is calculated as follows:
\begin{gather*}
S = 2p\begin{pmatrix}
\cos^2 \theta & 0 \\
0 & \sin^2 \theta
\end{pmatrix}, \quad S^{-1 / 2} = \begin{pmatrix}
\frac{1}{\sqrt{2p} \cos\theta} & 0 \\
0 & \frac{1}{\sqrt{2p} \sin\theta}
\end{pmatrix}, \\
E_1 = \begin{pmatrix}
1 / 2 & 1 / 2 \\
1 / 2 & 1 / 2
\end{pmatrix}, \quad E_2 = \begin{pmatrix}
1 / 2 & -1 / 2 \\
-1 / 2 & 1 / 2
\end{pmatrix}, \\
P_\mathcal{E}^{\text{PGM}+} = p\text{tr}(\sigma_1 E_1) + p\text{tr}(\sigma_2 E_2) = p(\cos\theta + \sin\theta)^2.
\end{gather*}
From Eq.~\eqref{equation:pgm_square_bound_tighter}, we have
\begin{align}
\label{equation:mirror_bound_naive}
P_\mathcal{E}^\text{opt} & \leq \sqrt{\frac{2}{3}\left( P_\mathcal{E}^{\text{PGM}} - \frac{1}{3} \right)} + \frac{1}{3} \nonumber \\
& = \sqrt{\frac{2}{3}\left( 2p \left( \frac{\sqrt{p} \cos^2 \theta}{\sqrt{1 - 2p \sin^2 \theta}} + \frac{\sin\theta}{\sqrt{2}} \right)^2 + \frac{(1 - 2p)^2}{1 - 2p\sin^2 \theta} - \frac{1}{3} \right)} + \frac{1}{3},
\end{align}
and from Eq.~\eqref{equation:pgm_improved_bound}, we have
\begin{align}
\label{equation:mirror_bound_improved}
P_\mathcal{E}^\text{opt} & \leq 2p \left\{ \sqrt{\frac{1}{2} \left( \frac{P_\mathcal{E}^{\text{PGM}+}}{2p} - \frac{1}{2} \right)} + \frac{1}{2} \right\} \nonumber \\
& = \sqrt{p P_\mathcal{E}^{\text{PGM}+} - p^2} + p \\
& = p (\sqrt{2 \cos\theta \sin\theta} + 1). \nonumber
\end{align}
One might expect that the bound in Eq.~\eqref{equation:mirror_bound_improved}, which exploits the additional information $\text{I}_+(\mathcal{E}) = \{ 1, 2 \}$, is tighter than the bound in Eq.~\eqref{equation:mirror_bound_naive} from naive PGM:
\begin{align}
\label{equation:desired_inequality}
\sqrt{\frac{2}{3}\left( 2p \left( \frac{\sqrt{p} \cos^2 \theta}{\sqrt{1 - 2p \sin^2 \theta}} + \frac{\sin\theta}{\sqrt{2}} \right)^2 + \frac{(1 - 2p)^2}{1 - 2p\sin^2 \theta} - \frac{1}{3} \right)} + \frac{1}{3} \nonumber \\
\geq p (\sqrt{2 \cos\theta \sin\theta} + 1).
\end{align}
Although not of general significance, it can be shown that Eq.~\eqref{equation:desired_inequality} holds for the following cases (Appendix~\ref{section:proof_special_cases}):
\begin{itemize}
\item $\theta = \pi / 4$ or $| \psi_1 \rangle \perp | \psi_2 \rangle$.
\item $p = 1 / 3$ or equiprobable states.
\end{itemize}
While Eq.~\eqref{equation:desired_inequality} is true for most pairs of $(\theta, p)$, for a small (seemingly convex) region, the inequality is in the opposite direction (see Fig.~\ref{fig:mirror_improve_region}). This nontrivial observation—that the bound applied to $\text{I}_+(\mathcal{E})$ is not always tighter—motivates further research to identify the conditions under which the bound tightens.

\subsubsection{\hl{Random ensembles}}

\begin{table}[h]
    \centering
    \renewcommand{\arraystretch}{1.5}
    \caption{\hl{For each pair $(n,N)$, we generated $100$ random ensembles of $N$ states on $n$ qubits, separately for pure-state ensembles and mixed-state ensembles, and compared Eq.~\eqref{equation:pgm_square_bound_tighter} with Eq.~\eqref{equation:pgm_improved_bound}. Each entry $a/b$ gives the fraction of the $100$ pure-state instances (left) and the $100$ mixed-state instances (right) for which Eq.~\eqref{equation:pgm_improved_bound} is tighter than Eq.~\eqref{equation:pgm_square_bound_tighter}. The data show that although Eq.~\eqref{equation:pgm_improved_bound} is not uniformly tighter in low-dimensional cases, its advantage becomes increasingly prevalent as either the number of qubits $n$ or the ensemble size $N$ grows.}}
    \label{tab:random_ensembles}
    \begin{tabular}{|c|c|c|c|c|c|c|c|c|}
    \hline
    \diagbox{$n$}{$N$} & 3 & 4 & 5 & 6 & 7 & 8 & 9 & 10\\
    \hline
    1 & .86/.96 & .93/1.0 & .98/.99 & .99/1.0 & 1.0/1.0 & 1.0/1.0 & 1.0/1.0 & 1.0/1.0\\
    \hline
    2 & 1.0/1.0 & 1.0/1.0 & .80/1.0 & .94/1.0 & .96/1.0 & .99/1.0 & .99/1.0 & 1.0/1.0\\
    \hline
    3 & 1.0/1.0 & 1.0/1.0 & 1.0/1.0 & 1.0/1.0 & 1.0/1.0 & 1.0/1.0 & .72/1.0 & .83/1.0\\
    \hline
    4 & 1.0/1.0 & 1.0/1.0 & 1.0/1.0 & 1.0/1.0 & 1.0/1.0 & 1.0/1.0 & 1.0/1.0 & 1.0/1.0\\
    \hline
    5 & 1.0/1.0 & 1.0/1.0 & 1.0/1.0 & 1.0/1.0 & 1.0/1.0 & 1.0/1.0 & 1.0/1.0 & 1.0/1.0\\
    \hline
    \end{tabular}
\end{table}

\hl{
To assess more broadly how Eq.~\eqref{equation:pgm_improved_bound} compares with Eq.~\eqref{equation:pgm_square_bound_tighter}, we performed additional numerical experiments beyond the single-qubit examples discussed above. Specifically, for each $(n,N)$ with $n=1,2,3,4,5$ and $N=3,4,5,6,7,8,9,10$, we generated random ensembles of $N$ states on $n$ qubits and evaluated both bounds on each instance. We carried out this test separately for pure-state ensembles and mixed-state ensembles, using $100$ random samples in each class.

The results are summarized in Table~\ref{tab:random_ensembles}. They confirm that Eq.~\eqref{equation:pgm_improved_bound} is not uniformly tighter than Eq.~\eqref{equation:pgm_square_bound_tighter}. Even in the case of three single-qubit states, one finds nontrivial fractions of random instances for which Eq.~\eqref{equation:pgm_improved_bound} does not improve upon Eq.~\eqref{equation:pgm_square_bound_tighter}. Thus, the superiority of Eq.~\eqref{equation:pgm_improved_bound} cannot be claimed in full generality.

At the same time, the table also reveals a clear empirical pattern. As either the number of qubits $n$ or the number of states $N$ increases, the proportion of tested instances in which Eq.~\eqref{equation:pgm_improved_bound} is tighter becomes very high, and in many of the larger configurations it is close to $1.0$ for both pure and mixed ensembles.
}

\subsubsection{Equiprobable states}
For equiprobable states, the inequality that Eq.~\eqref{equation:pgm_improved_bound} is less than or equal to Eq.~\eqref{equation:pgm_square_bound_tighter} reduces to:
\begin{align*}
& \frac{|\text{I}_+(\mathcal{E})|}{N} \left\{ \sqrt{\frac{|\text{I}_+(\mathcal{E})| - 1}{|\text{I}_+(\mathcal{E})|}\left( \frac{N}{|\text{I}_+(\mathcal{E})|} P_\mathcal{E}^{\text{PGM}+} - \frac{1}{|\text{I}_+(\mathcal{E})|} \right)} + \frac{1}{|\text{I}_+(\mathcal{E})|} \right\} \\
& \displaybreak[1] \\
= & \sqrt{\frac{|\text{I}_+(\mathcal{E})| - 1}{N} \left( P_\mathcal{E}^{\text{PGM}+} - \frac{1}{N} \right)} + \frac{1}{N} \\
\leq & \sqrt{\frac{N - 1}{N}\left( P_\mathcal{E}^{\text{PGM}} - \frac{1}{N} \right)} + \frac{1}{N},
\end{align*}
which is equivalent to
\begin{eqnarray}
\label{equation:equiprobable_pgm_bound_inequality}
(|\text{I}_+(\mathcal{E})| - 1) \left( P_\mathcal{E}^{\text{PGM}+} - \frac{1}{N} \right) \leq (N - 1) \left( P_\mathcal{E}^{\text{PGM}} - \frac{1}{N} \right).
\end{eqnarray}
Numerical attempts have failed to identify any exceptions to Eq.~\eqref{equation:equiprobable_pgm_bound_inequality}. Consequently, we conjecture that it holds universally for equiprobable states.

\begin{figure}[t]
\centering
\includegraphics[width=0.8\textwidth]{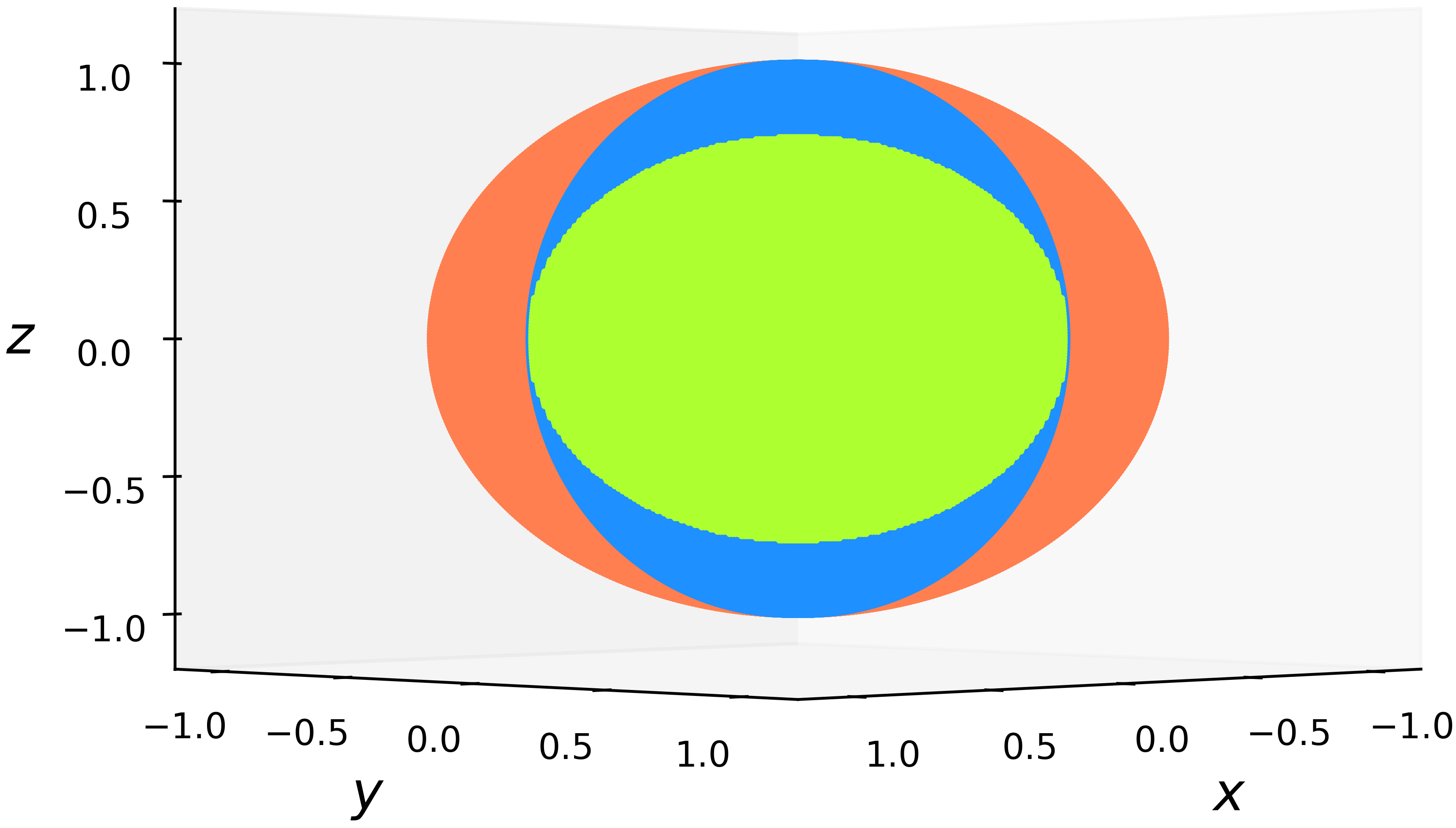}
\caption{\label{fig:conjecture_bloch} \hl{A geometric illustration for the simplest nontrivial instance of the equiprobable-state conjecture regarding Eq.~\eqref{equation:equiprobable_pgm_bound_inequality}.} $|\psi_1\rangle$ and $|\psi_2\rangle$ correspond to the Bloch vectors $(1, 0, 0)$ and $(0, 1, 0)$, respectively. (Red+Blue+Green) Bloch sphere for 2-level system. (Blue+Green) Bloch vectors corresponding to $|\psi_3\rangle$ for which $| \langle \psi_1 | \psi_2 \rangle | = \min\limits_{i, j} | \langle \psi_i | \psi_j \rangle |$. (Green) Bloch vectors corresponding to $|\psi_3\rangle$ for which $\text{I}_+(\mathcal{E}) = \{ 1, 2 \}$. It is numerically observed that all points in this region satisfy Eq.~(\ref{equation:inequality_pgm_dN23}).}
\end{figure}

\hl{The simplest nontrivial configuration for demonstrating this conjecture for equiprobable states involves pure states with $(d, N) = (2, 3)$ and $|\text{I}_+(\mathcal{E})| = 2$.}
For equiprobable states $\mathcal{E} = \big\{ | \Tilde{\psi}_1 \rangle \langle \Tilde{\psi}_1 |, | \Tilde{\psi}_2 \rangle \langle \Tilde{\psi}_2 |, | \Tilde{\psi}_3 \rangle \langle \Tilde{\psi}_3 | \big\}$, without loss of generality, let $\text{I}_+(\mathcal{E}) = \{ 1, 2 \}$. Then $| \langle \psi_1 | \psi_2 \rangle | = \min\limits_{i, j} | \langle \psi_i | \psi_j \rangle |$. This follows from the Helstrom bound \citep{barnett1997experimental}:
\begin{align*}
P_{\text{error}} & = \frac{1}{2} \left\{ 1 - \left[ 1 - 4 p_1 p_2 | \langle \psi_1 | \psi_2 \rangle |^2 \right]^{1 / 2} \right\}, \\
P_{\text{success}} & = \frac{1}{2} \left\{ 1 + \left[ 1 - 4 p_1 p_2 | \langle \psi_1 | \psi_2 \rangle |^2 \right]^{1 / 2} \right\},
\end{align*}
\hl{where the probabilities $p_1$ and $p_2=1-p_1$ correspond to $|\psi_1\rangle$ and $|\psi_2\rangle$, respectively.}
The optimal discrimination of two pure states $\Tilde{\mathcal{E}} = \left\{ \frac{| \psi_1 \rangle \langle \psi_1 |}{2}, \frac{| \psi_2 \rangle \langle \psi_2 |}{2} \right\}$ is the Helstrom measurement, which coincides with PGM \citep{chefles2000quantum}:
\begin{align*}
P_{\Tilde{\mathcal{E}}}^\text{opt} & = P_{\Tilde{\mathcal{E}}}^{\text{PGM}} \\
& = \frac{1}{2} \left\{ 1 + \left[ 1 - 4 \left( \frac{1}{3 p_+(\mathcal{E})} \right)^2 | \langle \psi_1 | \psi_2 \rangle |^2 \right]^{1 / 2} \right\} \\
& = \frac{1}{2} \left\{ 1 + \left[ 1 - | \langle \psi_1 | \psi_2 \rangle |^2 \right]^{1 / 2} \right\}.
\end{align*}
From Eq.~(\ref{equation:pgm_improved_bound}), we get
\begin{align*}
P_\mathcal{E}^\text{opt} & \leq \frac{2}{3} \left\{ \sqrt{\frac{1}{2}\left( P_{\Tilde{\mathcal{E}}}^{\text{PGM}} - \frac{1}{2} \right)} + \frac{1}{2} \right\} \\
& = \frac{2}{3} \left\{ \sqrt{\frac{\left[ 1 - | \langle \psi_1 | \psi_2 \rangle |^2 \right]^{1 / 2}}{4}} + \frac{1}{2} \right\} \\
& = \frac{1 + \sqrt[4]{1 - | \langle \psi_1 | \psi_2 \rangle |^2}}{3}.
\end{align*}
Then $(\ref{equation:pgm_improved_bound}) \leq (\ref{equation:pgm_square_bound_tighter})$ is equivalent to
\begin{align}
\label{equation:inequality_pgm_dN23}
& \sqrt{\frac{2}{3}\left( P_\mathcal{E}^{\text{PGM}} - \frac{1}{3} \right)} + \frac{1}{3} \geq \frac{1 + \sqrt[4]{1 - | \langle \psi_1 | \psi_2 \rangle |^2}}{3} \nonumber \\
\Leftrightarrow \quad & P_\mathcal{E}^{\text{PGM}} \geq \frac{\sqrt{1 - | \langle \psi_1 | \psi_2 \rangle |^2} + 2}{6}.
\end{align}
Fig.~\ref{fig:conjecture_bloch} illustrates an example.

\subsection{Other bounds}
In this section, three additional upper bounds on $P_\mathcal{E}^\text{opt}$ are considered \citep{ogawa1999strong, montanaro2008lower, tyson2009two, loubenets2022general}:
\begin{align}
P_\mathcal{E}^\text{opt} & \leq 1 - \sum_{1 \leq i < j \leq N} F_{ij}^2, \label{eq:other_bound_1} \\
P_\mathcal{E}^\text{opt} & \leq \text{tr} \left( \sqrt{\sum_{i = 1}^N \Tilde{\sigma}_i^2} \right), \label{eq:other_bound_2} \\
P_\mathcal{E}^\text{opt} & \leq \frac{1}{2} + \frac{1}{2} \min_{j = 1, \ldots , N} \left\{ \sum_{i = 1}^N \| \Tilde{\sigma}_i - \Tilde{\sigma}_j \|_1 - p_j (N - 2) \right\}, \label{eq:other_bound_3}
\end{align}
where $\| A \|_1 \equiv \text{tr} (\sqrt{A^\dagger A})$ denotes the \textit{trace norm} of $A$ \citep{fan1951maximum}.

\subsubsection{The bound in Eq.~\eqref{eq:other_bound_1} for mirror-symmetric states}
\label{sec:mirror_bound_2}

Consider the mirror-symmetric states in Eq.~\eqref{equation:mirror_definition}. Note that $F_{12} = p |\cos 2\theta|$ and $F_{13} = F_{23} = \sqrt{p(1 - 2p)} \cos\theta$. Then Eq.~\eqref{eq:other_bound_1} becomes
\begin{eqnarray}
\label{equation:mirror_fidelity_full}
P_\mathcal{E}^\text{opt} \leq 1 - p^2 \cos^2 2\theta - 2p(1 - 2p)\cos^2 \theta.
\end{eqnarray}
Suppose $|\text{I}_+(\mathcal{E})| = | \{ 1, 2 \} | = 2$. Applying Eq.~\eqref{eq:other_bound_1} to $\Tilde{\mathcal{E}}$ and using Eq.~\eqref{equation:pruned_relation}, we get
\begin{eqnarray}
\label{equation:mirror_fidelity_partial}
P_\mathcal{E}^\text{opt} \leq 2p \left( 1 - \frac{1}{4} \cos^2 2\theta \right).
\end{eqnarray}
But
\begin{align*}
& 1 - p^2 \cos^2 2\theta - 2p(1 - 2p)\cos^2 \theta - 2p \left( 1 - \frac{1}{4} \cos^2 2\theta \right) \\
= & (1 - 2p) \left\{ 
1 + 2p \left( \frac{1}{4} \cos^2 2\theta - \cos^2 \theta \right) \right\} \\
\geq & (1 - 2p)^2 \\
\geq & 0,
\end{align*}
so the bound in Eq.~\eqref{equation:mirror_fidelity_partial} is tighter than Eq.~\eqref{equation:mirror_fidelity_full}.

\subsubsection{The bound in Eq.~\eqref{eq:other_bound_1} for equiprobable states}
Assume that the states are equiprobable. Then Eq.~\eqref{eq:other_bound_1} becomes
\begin{eqnarray}
\label{eq:other_bound_1_normalized}
P_\mathcal{E}^\text{opt} \leq 1 - \sum_{1 \leq i < j \leq N} p_i p_j \hat{F}_{ij}^2 = 1 - \frac{1}{N^2} \sum_{1 \leq i < j \leq N} \hat{F}_{ij}^2.
\end{eqnarray}
Applying Eq.~\eqref{eq:other_bound_1_normalized} to $\Tilde{\mathcal{E}}$ and using Eq.~\eqref{equation:pruned_relation}, we get
\begin{align}
\label{equation:fidelity_bound_improved}
P_\mathcal{E}^\text{opt} & \leq \frac{|\text{I}_+(\mathcal{E})|}{N} \left( 
1 - \frac{1}{|\text{I}_+(\mathcal{E})|^2} \sum_{\substack{i, j \in \text{I}_+(\mathcal{E}), \\ i < j}} \hat{F}_{ij}^2 \right) \nonumber \\
& = \frac{|\text{I}_+(\mathcal{E})|}{N} - \frac{1}{N |\text{I}_+(\mathcal{E})|} \sum_{\substack{i, j \in \text{I}_+(\mathcal{E}), \\ i < j}} \hat{F}_{ij}^2.
\end{align}
The bound in Eq.~\eqref{equation:fidelity_bound_improved} is tighter than Eq.~\eqref{eq:other_bound_1_normalized}. To see this, note that $N \geq |\text{I}_+(\mathcal{E})|$. This gives
\begin{align*}
& (N - |\text{I}_+(\mathcal{E})|) (N - |\text{I}_+(\mathcal{E})| + 1) \geq 0 \\
\Rightarrow \quad & N^2 - 2N|\text{I}_+(\mathcal{E})| + N + |\text{I}_+(\mathcal{E})|^2 - |\text{I}_+(\mathcal{E})| \geq 0 \\
\Rightarrow \quad & 2N^2 - 2N|\text{I}_+(\mathcal{E})| \geq N^2 - N - |\text{I}_+(\mathcal{E})|^2 + |\text{I}_+(\mathcal{E})| \\
\Rightarrow \quad & 1 - \frac{|\text{I}_+(\mathcal{E})|}{N} \geq \frac{1}{N^2} \left( 
\frac{N(N - 1)}{2} - \frac{|\text{I}_+(\mathcal{E})| (|\text{I}_+(\mathcal{E})| - 1)}{2} \right).
\end{align*}
Now consider the following sum:
\begin{eqnarray*}
\sum_{\substack{\{i, j\} \not\subseteq \text{I}_+(\mathcal{E}), \\ i < j}} \hat{F}_{ij}^2 = \sum_{1 \leq i < j \leq N} \hat{F}_{ij}^2 - \sum_{\substack{i, j \in \text{I}_+(\mathcal{E}), \\ i < j}} \hat{F}_{ij}^2.
\end{eqnarray*}
The left-hand side consists of $\frac{N(N - 1)}{2} - \frac{|\text{I}_+(\mathcal{E})| (|\text{I}_+(\mathcal{E})| - 1)}{2}$ summands, all of which are less than or equal to $1$. Therefore,
\begin{align*}
& 1 - \frac{|\text{I}_+(\mathcal{E})|}{N} \geq \frac{1}{N^2} \sum_{\substack{\{i, j\} \not\subseteq \text{I}_+(\mathcal{E}), \\ i < j}} \hat{F}_{ij}^2 = \frac{1}{N^2} \left( \sum_{1 \leq i < j \leq N} \hat{F}_{ij}^2 - \sum_{\substack{i, j \in \text{I}_+(\mathcal{E}), \\ i < j}} \hat{F}_{ij}^2 \right) \\
\Rightarrow \quad & 1 - \frac{|\text{I}_+(\mathcal{E})|}{N} \geq \frac{1}{N^2} \sum_{1 \leq i < j \leq N} \hat{F}_{ij}^2 - \frac{1}{N|\text{I}_+(\mathcal{E})|} \sum_{\substack{i, j \in \text{I}_+(\mathcal{E}), \\ i < j}} \hat{F}_{ij}^2 \quad (\because N \geq |\text{I}_+(\mathcal{E})|) \\
\Rightarrow \quad & 1 - \frac{1}{N^2} \sum_{1 \leq i < j \leq N} \hat{F}_{ij}^2 \geq \frac{|\text{I}_+(\mathcal{E})|}{N} - \frac{1}{N|\text{I}_+(\mathcal{E})|} \sum_{\substack{i, j \in \text{I}_+(\mathcal{E}), \\ i < j}} \hat{F}_{ij}^2,
\end{align*}
as desired.

\begin{remark}
\hl{The case $|\text{I}_+(\mathcal{E})|<N$ does not contradict the geometrically uniform example in Section~\ref{sec:sufficiency}. In that case the states are not only equiprobable but also symmetry-related, so the PGM is optimal and all measurement outcomes are used on equal footing. However, we assume only equal priors while the geometry of the states remains otherwise unrestricted. Equal priors alone do not imply $|\text{I}_+(\mathcal{E})|=N$. When one state is sufficiently redundant relative to the others, assigning a POVM element to it can reduce the average success probability, and the optimal POVM may instead set that element to zero. This phenomenon already appears for three equally likely mirror-symmetric states in Eq.~\eqref{equation:mirror_definition} for an appropriate range of $\theta$.}
\end{remark}

\subsubsection{The bound in Eq.~\eqref{eq:other_bound_2} for arbitrary states}
It is rather straightforward to show that we always get a tighter bound on $P_\mathcal{E}^\text{opt}$ for arbitrary ensembles given $\text{I}_+(\mathcal{E})$. Applying Eq.~\eqref{eq:other_bound_2} to $\Tilde{\mathcal{E}}$ and using Eq.~\eqref{equation:pruned_relation}, we get
\begin{align}
\label{equation:psd_sum_improved_bound}
P_\mathcal{E}^\text{opt} & \leq p_+(\mathcal{E}) \text{tr} \left( \sqrt{\sum_{i \in \text{I}_+(\mathcal{E})} \frac{\Tilde{\sigma}_i^2}{p_+(\mathcal{E})^2}} \right) \nonumber \\
& = \text{tr} \left( \sqrt{\sum_{i \in \text{I}_+(\mathcal{E})} \Tilde{\sigma}_i^2} \right) \\
& \leq \text{tr} \left( \sqrt{\sum_{i = 1}^N \Tilde{\sigma}_i^2} \right), \nonumber
\end{align}
where the last inequality follows from the well-known \textit{L\"{o}wner-Heinz inequality,} specifically the operator monotonicity of the square root function \citep{lowner1934monotone}.

\subsubsection{The bound in Eq.~\eqref{eq:other_bound_3} for equiprobable states}
Assume that the states are equiprobable. Then Eq.~\eqref{eq:other_bound_3} becomes
\begin{align}
\label{eq:other_bound_3_equiprobable}
P_\mathcal{E}^\text{opt} & \leq \frac{1}{2} + \frac{1}{2} \min_{j = 1, \ldots , N} \left\{ \sum_{i = 1}^N \left\| \frac{\sigma_i}{N} - \frac{\sigma_j}{N} \right\|_1 - \frac{N - 2}{N} \right\} \nonumber \\
& = \frac{1}{2N} \left[ N + \min_{j = 1, \ldots , N} \left\{ \sum_{i = 1}^N \| \sigma_i - \sigma_j \|_1 \right\} - (N - 2) \right] \\
& = \frac{1}{2N} \left[ \min_{j = 1, \ldots , N} \left\{ \sum_{i = 1}^N \| \sigma_i - \sigma_j \|_1 \right\} + 2 \right]. \nonumber
\end{align}
Knowledge of the bound in Eq.~\eqref{eq:other_bound_3_equiprobable} typically implies that $\hat{j} \equiv \arg\min_{j = 1, \ldots , N} \left\{ \sum_{i = 1}^N \| \sigma_i - \sigma_j \|_1 \right\}$ is known. Applying Eq.~\eqref{eq:other_bound_3_equiprobable} to the ensemble $\left\{ \frac{N \Tilde{\sigma}_i}{\left| \text{I}_+(\mathcal{E}) \cup \left\{ 
\hat{j} \right\} \right|} \mid i \in \text{I}_+(\mathcal{E}) \cup \big\{ 
\hat{j} \big\} \right\}$ and using Eq.~\eqref{equation:pruned_relation}, we get
\begin{align}
\label{equation:trace_norm_improved_bound}
P_\mathcal{E}^\text{opt} & \leq \frac{ \left| \text{I}_+(\mathcal{E}) \cup \big\{ \hat{j} \big\} \right| }{N} \frac{1}{2 \left| \text{I}_+(\mathcal{E}) \cup \big\{ \hat{j} \big\} \right|} \left[ \min_{j \in \text{I}_+(\mathcal{E}) \cup \left\{ \hat{j} \right\} } \left\{ \sum_{i \in \text{I}_+(\mathcal{E}) \cup \left\{ \hat{j} \right\} } \| \sigma_i - \sigma_j \|_1 \right\} + 2 \right] \nonumber \\
& = \frac{1}{2N} \left[ \min_{j \in \text{I}_+(\mathcal{E}) \cup \left\{ \hat{j} \right\} } \left\{ \sum_{i \in \text{I}_+(\mathcal{E}) \cup \left\{ \hat{j} \right\} } \| \sigma_i - \sigma_j \|_1 \right\} + 2 \right].
\end{align}
But since
\begin{eqnarray*}
\sum_{i = 1}^N \| \sigma_i - \sigma_{\hat{j}} \|_1 \geq \sum_{i \in \text{I}_+(\mathcal{E}) \cup \left\{ \hat{j} \right\} } \| \sigma_i - \sigma_{\hat{j}} \|_1 ,
\end{eqnarray*}
the bound in Eq.~\eqref{equation:trace_norm_improved_bound} is tighter than Eq.~\eqref{eq:other_bound_3_equiprobable}.

\section{Derivation of subsets and supersets of $\text{I}_+(\mathcal{E})$}
\label{section:subset_superset}
\hl{Geometric analyses of minimum-error discrimination in which vanishing optimal POVM elements play an important role appear in \cite{ha2013complete, rouhbakhsh2023geometric}. Likewise, the results from Section~\ref{section:sparsity} can provide partial information about $\text{I}_+(\mathcal{E})$.} That is, it can be determined that $i \in \text{I}_+(\mathcal{E})$ for some $i$. To see this, we consider a simple example. The lower bound on $P_\mathcal{E}^\text{opt}$ closely related to Eq.~\eqref{eq:other_bound_2} is
\begin{eqnarray*}
P_\mathcal{E}^\text{opt} \geq \left[ \text{tr} \left( \sqrt{\sum_{i = 1}^N \Tilde{\sigma}_i^2} \right) \right]^2
\end{eqnarray*}
\citep{tyson2009two}. Combining this inequality with Eq.~\eqref{equation:psd_sum_improved_bound} we get
\begin{eqnarray}
\label{equation:lb_up_inequality}
\text{tr} \left( \sqrt{\sum_{i \in \text{I}_+(\mathcal{E})} \Tilde{\sigma}_i^2} \right) \geq \left[ \text{tr} \left( \sqrt{\sum_{i = 1}^N \Tilde{\sigma}_i^2} \right) \right]^2.
\end{eqnarray}
Since the left-hand side is monotonic in $\text{I}_+(\mathcal{E})$, we conclude that if replacing $\text{I}_+(\mathcal{E})$ with $\{ 1, 2, \ldots , N \} \setminus \{ i \}$ on the left-hand side of Eq.~\eqref{equation:lb_up_inequality} violates the inequality, then $i \in \text{I}_+(\mathcal{E})$. Clearly, this method does not allow us to uniquely determine $\text{I}_+(\mathcal{E})$: it only finds a subset of $\text{I}_+(\mathcal{E})$.

Another way to find a subset of $\text{I}_+(\mathcal{E})$ is to form \(A_{ij}=\tilde{\sigma}_i-\tilde{\sigma}_j\) and its positive part \(A_{ij}^+\) for each \(i\ne j\). Define
\[
\Pi_i = \bigcap_{j\ne i} \text{supp}(A_{ij}^+).
\]
If \(\text{dim}(\Pi_i)>0\), then \(i\in\text{I}_+(\mathcal{E})\). This is because for \(|\psi\rangle\in\Pi_i\) we have \(\langle\psi|\tilde{\sigma}_i|\psi\rangle > \langle\psi|\tilde{\sigma}_j|\psi\rangle\) for every \(j\ne i\). Therefore, the orthogonal projection on \(\Pi_i\) must be contained in \(\hat{E}_i (\mathcal{E})\).

We can also find a superset of $\text{I}_+(\mathcal{E})$ by using a sufficient condition for $\hat{E}_i (\mathcal{E}) = 0$. Suppose there exists a set of nonnegative coefficients $\{ \omega_i \}_{i = 1}^{N - 1}$ such that $\sum_{i = 1}^{N - 1} \omega_i = 1$ and
\begin{eqnarray*}
\sum_{j = 1}^{i - 1} \omega_j \Tilde{\sigma}_j + \sum_{j = i + 1}^{N} \omega_{j - 1} \Tilde{\sigma}_j \succ \Tilde{\sigma}_i.
\end{eqnarray*}
Then the $E = \{ E_j \}_{j = 1}^N$ defined as
\begin{eqnarray*}
E_j \equiv \begin{cases}
\hat{E}_j (\mathcal{E}) + \omega_j \hat{E}_i (\mathcal{E}) & \text{if } j < i \\
0 & \text{if } j = i \\
\hat{E}_j (\mathcal{E}) + \omega_{j - 1} \hat{E}_i (\mathcal{E}) & \text{if } j > i
\end{cases}
\end{eqnarray*}
is a valid POVM as it satisfies $E_j \succeq 0$ and $\sum_{j = 1}^{N} E_j = I$. Also,
\begin{align*}
P_{\mathcal{E}, E}^\text{success} & = \sum_{j \neq i} \text{tr} \left( \Tilde{\sigma}_j \hat{E}_j (\mathcal{E}) \right) + \text{tr} \left( \left( \sum_{j = 1}^{i - 1} \omega_j \Tilde{\sigma}_j + \sum_{j = i + 1}^{N} \omega_{j - 1} \Tilde{\sigma}_j \right) \hat{E}_i (\mathcal{E}) \right) \\
& \geq \sum_{j \neq i} \text{tr} \left( \Tilde{\sigma}_j \hat{E}_j (\mathcal{E}) \right) + \text{tr} \left( \Tilde{\sigma}_i \hat{E}_i (\mathcal{E}) \right) \\
& = P_\mathcal{E}^\text{opt},
\end{align*}
where the equality holds if and only if $\hat{E}_i (\mathcal{E}) = 0$. But since $P_{\mathcal{E}, E}^\text{success} \leq P_\mathcal{E}^\text{opt}$, the equality must hold and therefore $\hat{E}_i (\mathcal{E}) = 0$.

\begin{figure}[t]
\centering
\includegraphics[width=0.5\textwidth]{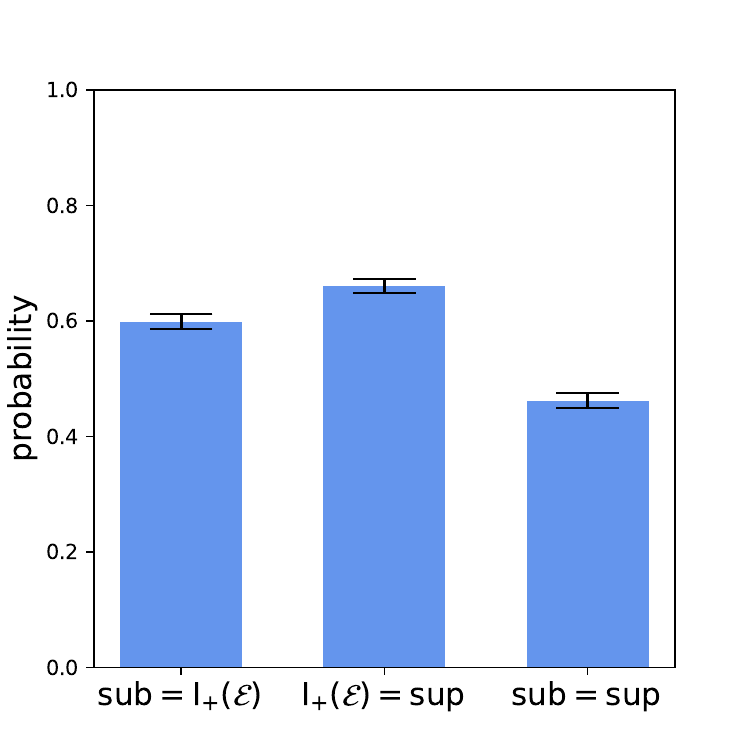}
\caption{\label{fig:subset_truth_superset} Probabilities of coincidence for the subsets and supersets found as described in Section~\ref{section:subset_superset}, with 99\% confidence intervals.}
\end{figure}

These methods were evaluated using three single-qubit density matrices independently sampled according to the Hilbert-Schmidt metric \citep{zyczkowski2001induced}. Probability distributions were sampled from the Dirichlet distribution \( \text{Dir}(1, 1, 1) \). For superset identification, the weights were set as \(\omega_i = \frac{1}{N - 1}\). 10,000 instances of such QSD problems were generated. Fig.~\ref{fig:subset_truth_superset} shows the results. In approximately half of the cases, the subset and superset of \( \text{I}_{+}(\mathcal{E}) \) coincided, uniquely determining \( \text{I}_{+}(\mathcal{E}) \) without SDP.

In general, determining $\text{I}_+(\mathcal{E})$ more efficiently than solving a full SDP remains a challenge. The union of any collection of identified subsets of \(\text{I}_+(\mathcal{E})\) is a subset of \(\text{I}_+(\mathcal{E})\), and the intersection of any collection of identified supersets of \(\text{I}_+(\mathcal{E})\) is a superset of \(\text{I}_+(\mathcal{E})\). Therefore, finding better necessary or sufficient conditions for $\hat{E}_i (\mathcal{E}) \succ 0$ or $\hat{E}_i (\mathcal{E}) = 0$, either mathematical or heuristic, will improve the results.

\section{Conclusion}
\label{section:conclusion}
In this work, we have examined some structural information about quantum state discrimination. \hl{On the state side, we revisited the single-qubit pure-state setting, where pairwise fidelities determine the optimal discrimination probability.} In higher-dimensional systems, such a simple correspondence no longer holds. Inspired by this observation, we derived a closed-form expression for three equiprobable single-qubit states with equal pairwise fidelities. On the measurement side, we showed that knowing which measurement operators must vanish can lead to tighter bounds on the optimal discrimination probability. Moreover, we provide examples showing that subsets and supersets of nonvanishing operators can be inferred without SDP. The challenge of determining every nonvanishing operator without SDP—either analytically or through more efficient numerical methods—remains largely open. A deeper understanding of how to quickly identify this set could further enhance our ability to find the optimal POVM, as it would reduce the computational complexity required for solving the associated SDP.

\begin{appendices}

\section{Derivation of $\delta$}
\label{section:delta_derivation}
Note that
\begin{align*}
| \langle \psi_2 | \psi_3 \rangle |^2 & = |\alpha^2 + (1 - \alpha^2) e^{i\theta}|^2 \\
& = |\alpha^2 + (1 - \alpha^2) \cos\theta + i(1 - \alpha^2) \sin\theta|^2 \\
& = \alpha^4 + 2 \alpha^2 (1 - \alpha^2) \cos\theta + (1 - \alpha^2)^2 \cos^2 \theta + (1 - \alpha^2)^2 \sin^2 \theta \\
& = 2\alpha^4 - 2\alpha^2 + 1 + 2 \alpha^2 (1 - \alpha^2) \cos\theta \\
& = \alpha^2,
\end{align*}
so
\begin{eqnarray*}
\cos\theta = \frac{2\alpha^4 - 3\alpha^2 + 1}{2 \alpha^2 (\alpha^2 - 1)}.
\end{eqnarray*}
Therefore,
\begin{align*}
\delta & = AD - BC \\
& = 2(1 - \alpha^2)(1 + 2\alpha^2) - (1 + e^{i\theta}) (1 + e^{-i\theta}) \alpha^2 (1 - \alpha^2) \\
& = 2(1 - \alpha^2)(1 + 2\alpha^2) - 2 (1 + \cos \theta) \alpha^2 (1 - \alpha^2) \\
& = 2(1 - \alpha^2) \{ 1 + (1 - \cos \theta) \alpha^2 \} \\
& = 3(1 - \alpha^2).
\end{align*}

\section{Proof of Eq.~\eqref{equation:desired_inequality} for special cases}
\label{section:proof_special_cases}
\begin{itemize}
\item If $\theta = \pi / 4$, then
\begin{equation*}
\mathrm{Eq.}~\eqref{equation:desired_inequality} \quad \Leftrightarrow \quad \sqrt{\frac{2}{3}\left( 2p \left( \frac{\sqrt{p}}{2\sqrt{1 - p}} + \frac{1}{2} \right)^2 + \frac{(1 - 2p)^2}{1 - p} - \frac{1}{3} \right)} + \frac{1}{3} \nonumber \geq 2p,
\end{equation*}
where $1 / \{ 2 + \cos\theta (\cos\theta + \sin\theta) \} = 1 / 3 \leq p \leq 1 / 2$. Continuing,
\begin{align*}
\Leftrightarrow \quad & \frac{2}{3}\left( 2p \left( \frac{\sqrt{p}}{2\sqrt{1 - p}} + \frac{1}{2} \right)^2 + \frac{(1 - 2p)^2}{1 - p} - \frac{1}{3} \right) \nonumber \geq \left( 2p - \frac{1}{3} \right)^2 \\
\Leftrightarrow \quad & \frac{9p^2 - 8p + 2}{3(1 - p)} - 4p^2 + \frac{5p}{3} + \frac{2p\sqrt{p}}{3\sqrt{1 - p}} - \frac{1}{3} \geq 0.
\end{align*}
But note that
\begin{align*}
& (2p - 1)^2 (5p - 1) = 20p^3 - 24p^2 + 9p - 1 \geq 0 \\
\Leftrightarrow \quad & \frac{4p^3}{1 - p} \geq (4p - 1)^2 \\
\Leftrightarrow \quad & \frac{2p\sqrt{p}}{\sqrt{1 - p}} - 4p + 1 = 3 \left( \frac{2p\sqrt{p}}{3\sqrt{1 - p}} - \frac{1}{3} - \frac{4}{3} \left( p - \frac{1}{2} \right) \right) \geq 0 \\
\Leftrightarrow \quad & \frac{2p\sqrt{p}}{3\sqrt{1 - p}} - \frac{1}{3} \geq \frac{4}{3} \left( p - \frac{1}{2} \right),
\end{align*}
so
\begin{align*}
\frac{9p^2 - 8p + 2}{3(1 - p)} - 4p^2 + \frac{5p}{3} + \frac{2p\sqrt{p}}{3\sqrt{1 - p}} - \frac{1}{3} & \geq \frac{9p^2 - 8p + 2}{3(1 - p)} - 4p^2 + \frac{5p}{3} + \frac{4}{3} \left( p - \frac{1}{2} \right) \\
& = \frac{p(1 - 2p)^2}{1 - p} \\
& \geq 0.
\end{align*}
\item If $p = 1 / 3$, then
\begin{eqnarray*}
\mathrm{Eq.}~\eqref{equation:desired_inequality} \quad \Leftrightarrow \quad \sqrt{\frac{2}{3}\left( \frac{2}{3} \left( \frac{\cos^2 \theta}{\sqrt{3 - 2 \sin^2 \theta}} + \frac{\sin\theta}{\sqrt{2}} \right)^2 + \frac{1}{9 - 6\sin^2 \theta} - \frac{1}{3} \right)} + \frac{1}{3} \\
\geq \frac{\sqrt{2 \cos\theta \sin\theta} + 1}{3},
\end{eqnarray*}
where $0 \leq \theta \leq \pi / 4$. Continuing,
\begin{align*}
\Leftrightarrow \quad & 2 \left( \frac{\cos^2 \theta}{\sqrt{3 - 2\sin^2 \theta}} + \frac{\sin\theta}{\sqrt{2}} \right)^2 + \frac{1}{3 - 2\sin^2 \theta} - 1 \geq \cos\theta\sin\theta \\
\Leftarrow \quad & 2 \left( \frac{\cos^2 \theta}{\sqrt{3}} + \frac{\sin\theta}{\sqrt{2}} \right)^2 - \cos\theta\sin\theta - \frac{2}{3} \geq 0 \\
\Leftrightarrow \quad & 2 \left( \frac{1 - x^2}{\sqrt{3}} + \frac{x}{\sqrt{2}} \right)^2 - x\sqrt{1 - x^2} - \frac{2}{3} \geq 0, \quad 0 \leq x \leq \frac{1}{\sqrt{2}}.
\end{align*}
But since $\sqrt{1 - x^2} \leq 1 - x^2 / 2$,
\begin{eqnarray*}
\Leftarrow \quad f(x) = 2 \left( \frac{1 - x^2}{\sqrt{3}} + \frac{x}{\sqrt{2}} \right)^2 - x\left(1 - \frac{x^2}{2}\right) - \frac{2}{3} \geq 0.
\end{eqnarray*}
Note that
\begin{eqnarray*}
f\left(\frac{-1}{\sqrt{2}}\right) < 0, \quad f(0) = 0, \quad f\left(\frac{1}{\sqrt{2}}\right) > 0, \quad f(1) < 0.
\end{eqnarray*}
Since $f$ is a quartic polynomial with a positive leading coefficient, it has roots in
\begin{eqnarray*}
\left(-\infty, \frac{-1}{\sqrt{2}}\right), \quad 0, \quad \left( \frac{1}{\sqrt{2}}, 1 \right), \quad (1, \infty).
\end{eqnarray*}
Threrfore, $f(x) \geq 0$ for $0 \leq x \leq 1 / \sqrt{2}$.
\end{itemize}





\end{appendices}

\subsection*{Author contributions}
H.C. provided the main ideas, conducted the experiments, and wrote the manuscript. J.L. analyzed the results and revised the manuscript.

\subsection*{Data availability}
No datasets were generated or analyzed during the current study.

\section*{Declarations}
\subsection*{Competing interests}
The authors declare no competing interests.

\bibliography{sn-article}

\end{document}